# ‘The Moving Frontier’: How the National Justice Project Is Using Digital Technology to Expand Access to Justice

George Newhouse AM, Trisha Ramesh and Tiara de Silva[1]

## ABSTRACT

Over the past three decades, technological advances have assisted individuals to gain access to legal information, enhance the efficiency of legal services, and streamline dispute resolution. More recently, the proliferation of generative artificial intelligence has reinforced this emphasis on firm-level productivity, with commercial legal practice increasingly leveraging artificial intelligence tools to automate document creation and optimise case management. While commercial technology focuses on productivity, community organisations are exploring how artificial intelligence can level the playing field for marginalised communities. Far less attention has been given to free, direct-to-consumer legal technologies, as these models are often unprofitable. Consequently, justice technology scholarship has overlooked a distinct class of technology: digital reporting platforms that enable people to record their experiences of harm, while simultaneously aggregating those accounts into a body of evidence capable of supporting advocacy, institutional accountability, and law reform. Crucially, experiences of discrimination and administrative abuse are often dismissed by oversight bodies as individual grievances. To understand the transformative potential of these platforms, this article asks two questions. First, how can digital reporting close the knowledge and power gap for marginalised communities interacting with the legal system? Second, how can these tools collect evidence to hold institutions accountable, drive change and contribute to a more just society?

To address these questions, this article draws on a six-year case study of justice technology development at the National Justice Project (“NJP”), an Australian legal service dedicated to addressing systemic discrimination which was undertaken in conjunction with its pro-bono partners, Macquarie University, the University of New South Wales, the Jumbunna Institute for Indigenous Education (“the Jumbunna Institute”), the Kaldor Centre for International Refugee Law, and K&L Gates. The NJP’s journey reflects an evolution from early legal chatbots into direct-to-consumer reporting platforms, specifically *Call It Out* and *Hear Me Out*, which combine direct legal self-help with community data collection. This article argues that direct-to-consumer digital reporting platforms, by transforming individual experiences of harm into collective evidence, equip civil society with the evidence needed to self-advocate and challenge institutional discrimination. By using anonymised reporting, triage templates, and Retrieval-Augmented Generation (RAG) architectures, platforms like *Hear Me Out* and *Call It Out* record experiences of discrimination that other institutions rarely seek to collect. The aggregated data generated by the NJP’s platforms have directly informed parliamentary inquiries, supported annual analytical reports, and supplied data for royal commissions. By expanding the conceptualisation of justice technology beyond service delivery, this article demonstrates that documenting lived experience and producing evidence for social change are as fundamental to access to justice as providing legal advice.

## I INTRODUCTION

Individually reported experiences of discrimination, institutional misconduct, and administrative abuse are often dismissed by complaint bodies as isolated grievances. Traditional legal oversight architectures often place the burden on affected individuals to identify a legal problem, find the institution with jurisdiction to address it, and engage with the formal reporting process without assistance.[2] Consequently, administrative complaint data captures only a fraction of the harms experienced in the community.[3] For marginalised groups, this reliance on individual reporting creates a knowledge and power gap that, more broadly, prevents systemic patterns of harm less visible to legal institutions.[4] Over the past six years, the National Justice Project (NJP), an Australian human rights legal service specialising in strategic litigation and systemic advocacy, has developed a series of technology-based access-to-justice initiatives to respond to these barriers. The NJP's work has evolved from early legal chatbots into direct-to-consumer reporting platforms, including *Call It Out* and *Hear Me Out*. These platforms reduce the practical work required to identify a complaint pathway, formulate a complaint, and contribute to a wider account of harm. These technologies were designed with a dual purpose: first, to improve individual access to justice through automated triage and self-help; and second, to generate a body of empirical evidence capable of supporting systemic advocacy and law reform.[5] Drawing on scholarship across access to justice and data justice, this article examines two primary questions. First, how can digital reporting close the knowledge and power gap for marginalised communities interacting with the legal system? Second, how can these tools collect evidence to hold institutions accountable, drive change, and contribute to a more just society? Using a qualitative case-study methodology, this paper examines the evolution of these technologies, the theory of change underpinning them, and what they contribute to justice technology scholarship.

---

[1] George Newhouse is an Adjunct Professor of Law at Macquarie University and an Adjunct Professor with the Jumbunna Institute of Indigenous Education and Research at UTS and the CEO of the National Justice Project in Australia. Trisha Ramesh and Tiara de Silva are both law students with the National Justice Project. We also acknowledge the work of Professor Daniel Ghezelbash and Dean Moutopoulos in developing NJP’s AI tools.

[2] NSW Ombudsman, *Review of the DCJ Complaint System in respect of its Aboriginal Child Protection functions* (Special Report, 6 November 2024); Australian Human Rights Commission, *Inquiry into Freedom of Speech: Australian Human Rights Commission Submission to the Parliamentary Joint Committee on Human Rights* (Submission No 13, 9 December 2016).

## II THE RATIONALE AND THEORY OF CHANGE

Traditional legal service delivery cannot address the full extent of institutional harm experienced across the community. In Australia, unmet legal need is disproportionately experienced by marginalised populations. Many such harms are never addressed by a formal legal or institutional process, because people may lack legal knowledge, face prohibitive costs, or fail to recognise their experience as a legal harm. Further, 2024 data shows that 13.6% of the population living below the poverty line in Australia remain ineligible for public legal representation.[6] This access deficit is severely compounded by the fragmentation of Australia's administrative law landscape, encompassing nearly 300 federal, state and territory official complaint bodies.[7] For people without legal assistance, identifying the appropriate oversight body and preparing a complaint can itself be a barrier to accessing the complaints process, especially where complaint processes are procedurally opaque or culturally unsafe.[8] Formal complaint data therefore provides an incomplete basis for identifying recurring patterns of harm, particularly where people face barriers to making a complaint in the first instance.[9]

The NJP developed its technology projects as counter-hegemonic infrastructure to address these limitations, operationalising access to justice as a 'moving frontier' that extends justice-seeking beyond conventional courts and government oversight bodies into non-judicial avenues of collective evidence-building and advocacy.[10] The NJP's theory of change operates through four interconnected pillars.

1. First, as an access to justice tool, plain-language digital interfaces scaffold user interactions, which enable individuals to identify potential rights violations and reporting processes, and to take action, without requiring formal legal representation.
2. Second, direct reporting functions as an act of rights advocacy, by allowing users to describe experiences in their own terms.[11]
3. Third, aggregation operates as a form of data justice, by synthesising individual reports to reveal patterns of harm that may not be visible from individual complaints. For Call It Out, these decisions are informed by First Nations data sovereignty, with the Jumbunna Institute for Indigenous Education and Research at UTS ("the Jumbunna Institute"), controlling access to the data and First Nations researchers analysing reports to identify patterns of racism.[12]
4. Lastly, empirical visibility provides a foundation for institutional accountability. A body of complaints can provide evidence about the prevalence and context of harm and inform parliamentary inquiries, strategic litigation and law reform.[13]

---

[3] OECD, *Equal Access to Justice for Inclusive Growth: Putting People at the Centre* (OECD Publishing, 2019). See also Christine Coumarelos, Deborah Macourt, Julie People, Hugh M McDonald, Zhigang Wei, Reiny Iriana and Stephanie Ramsey, *Legal Australia-Wide Survey: Legal Need in Australia* (Law and Justice Foundation of New South Wales, 2012).

[4] Australian Human Rights Commission, *Inquiry into Freedom of Speech: Australian Human Rights Commission Submission to the Parliamentary Joint Committee on Human Rights* (Submission No 13, 9 December 2016).

[5] University of New South Wales, 'A New Era for Access to Justice: AI-Powered Platform "Hear Me Out" Launches in Australia' (24 March 2025); National Justice Project, 'Award-Winning NSW Pilot of AI Technology for Complaints Justice' (31 October 2024).

[6] Australian Council of Social Service and UNSW Sydney, *Poverty in Australia 2024* (2024).

[7] University of New South Wales, 'A New Era for Access to Justice: AI-Powered Platform "Hear Me Out" Launches in Australia' (24 March 2025).

[8] NSW Ombudsman, *Review of the DCJ Complaint System in respect of its Aboriginal Child Protection functions* (Special Report, 6 November 2024).

[9] Ibid.

[10] Marc Galanter, "Access to Justice as a Moving Frontier" in Julia Bass, W.A. Bogart & Frederick H. Zemans (eds), *Access to Justice for a New Century: The Way Forward* (Law Society of Upper Canada, 2005) 147–161.

[11] Patricia Ewick and Susan S Silbey, *The Common Place of Law: Stories from Everyday Life* (University of Chicago Press, 1998).

[12] Maiam nayri Wingara Indigenous Data Sovereignty Collective, 'MnW Principles' (Web Page). See also Linnet Taylor, 'What Is Data Justice? The Case for Connecting Digital Rights and Freedoms Globally' (2017) 4(2) *Big Data & Society* 1, 1–3.

[13] Marc Galanter, 'Why the "Haves" Come Out Ahead: Speculations on the Limits of Legal Change' (1974) 9(1) *Law & Society Review* 95, 95–160.

# III LESSONS FROM SIX YEARS OF JUSTICE TECHNOLOGY DEVELOPMENT

Justice technology evolves through iteration, with each platform emerging from the strengths and limitations of the model that preceded it. The NJP's digital strategy originated with CopWatch, which is a web application and community education program established in 2017 following reports to the NJP by First Nations families regarding pervasive police discrimination in Kalgoorlie following the death of Elijah Doughty.[14] The NJP implemented the CopWatch training and a local campaign in regional centres including Broken Hill and Dubbo to enable community members to safely document police interactions, creating a decentralised source of evidence to challenge discretionary policing practices, as well as training on how to de-escalate interactions.[15] Additionally, the local targeted media and advocacy campaigns around the CopWatch training helped to improve police-community relations.[16]

Building on this, the NJP's technology deployment has developed across three phases.

## 1 *Phase 1: Rules-Based Legal Chatbots and the Tech4Justice Clinic*

To assist people navigating complex discrimination complaints, the NJP partnered with Macquarie Law School in 2020 to establish the Tech4Justice Lab.[17] Tech4Justice operates as a student-led lab and clinic in which law students build AI-powered chatbots to assist people in making discrimination complaints.[18] The clinic's founders, supported by student feedback, distinguished this multi-year model from short-term, hackathon-style initiatives, which can often generate promising prototypes yet consistently struggle with long-term implementation.

Using an off-the-shelf no-code logic engine, students mapped complex statutory frameworks into conversational decision pathways designed to guide users through the law.[19] Through structured intake questions, the platform evaluated basic legal standing, jurisdictional requirements and appropriate complaint bodies.[20] However, codifying anti-discrimination law into pre-programmed decision pathways proved exceptionally difficult to scale. For example, mapping disability discrimination law required 268 conditional questions to account for intersecting variables, such as geographic jurisdiction, reasonable adjustment standards, and industry-specific regulations.[21] This hyper-rationalised architecture demonstrated the limits of rules-based systems, which fail to capture the unstructured, open-ended ways people articulate their personal experiences and trauma.

## 2 *Phase 2: Call It Out*

Recognising this limitation, the NJP subsequently developed a different model with Call It Out: a community governed platform designed to collect experiences of racism directly from First Nations communities. Co-developed with the Jumbunna Institute, Call It Out was launched on 21 March 2022 as an anonymised First Nations Racism Register.[22] Drawing on established community-led reporting initiatives, including the Islamophobia Register and similar reporting mechanisms developed by Jewish and Chinese communities in Australia, Call It Out was developed in response to the absence of a national community-based platform for reporting racism against First Nations people.[23] Unlike statutory complaint channels, Call It Out accepts reports from witnesses and bystanders alongside directly affected individuals, using anonymisation to protect reporting parties from institutional retaliation or re-traumatisation.[24] All submissions are owned and analysed by First Nations researchers at the Jumbunna Institute under First Nations Data Sovereignty principles.[25] The Register received 267 validated submissions during its first six months of operation in 2022, which provided an initial baseline for community-governed data collection.[26] Its first annual report, *In Every Corner of Every Suburb,* analysed 497 reports received between March 2022 and March 2023, while the 2023–2024 annual report analysed a further 453 validated reports.[27] These reports documented racism across workplaces, commercial settings, educational institutions, government institutions and public settings, providing

[14] Sofie Wainwright, 'Copwatch Program Empowers Aboriginal Communities to Film Interactions with Police', *ABC News* (online, 22 August 2017); Sarah Collard, 'Workshops Urge People to Film on Phones to Hold Them to Account', *ABC News* (online, 4 August 2018).

[15] National Justice Project, *2023 Impact Report* (Report, 2023) 23. See also Sofie Wainwright, 'Copwatch Program Empowers Aboriginal Communities to Film Interactions with Police', *ABC News* (online, 22 August 2017).

[16] Ibid.

[17] Danielle Moon, Heike Fabig, George Newhouse and Daniel Ghezelbash, 'Teaching Legal Design: Lessons from Five Years of Student-Led Innovation in Tech4Justice' (2025) 2(1) *Legal Design Journal* 1, 4.

[18] Ibid, 4–5.

[19] Ibid, 5.

[20] Ibid.

[21] Ibid.

[22] Jumbunna Institute for Indigenous Education and Research and National Justice Project, *Call It Out Racism Register: The First Six Months* (Report, 2022) 3; Jumbunna Institute for Indigenous Education and Research, 'Call It Out Racism Register', University of Technology Sydney (Web Page).

[23] Jumbunna Institute for Indigenous Education and Research, 'Call It Out Racism Register', University of Technology Sydney (Web Page)

[24] Jumbunna Institute for Indigenous Education and Research and National Justice Project, *Call It Out Racism Register: The First Six Months*, 4–6.

[25] Ibid.

[26] Ibid 34.

[27] Jumbunna Institute for Indigenous Education and Research and National Justice Project, *In Every Corner of Every Suburb: The Call It Out Racism Register 2022–2023* (Report, 2023) 3; Jumbunna Institute for Indigenous Education and Research and National Justice Project, *If You Don't Think Racism Exists Come Take a Walk with Us: Call It Out Annual Report 2023–2024* (Report, 2025).

evidence of the range of contexts in which First Nations people and witnesses reported experiencing racism.[28]

As the only national, Indigenous-led collection of data on racism against First Nations people, the Register's data has been used directly in advocacy and law reform. In 2026, Call It Out and the Jumbunna Institute contributed evidence to the Parliamentary Joint Committee on Aboriginal and Torres Strait Islander Affairs' *Inquiry into Racism, Hate and Violence directed at Aboriginal and Torres Strait Islander people* (the Racism Inquiry).[29] The inquiry heard evidence drawn from data collected through the Register, including evidence about the prevalence and contexts of racism experienced by First Nations people.[30] The evidence provided a basis for examining the limitations of formal reporting mechanisms and the extent of underreporting in official complaint data, often resulting from a complete breakdown of community trust. Further, during the Voice to Parliament referendum period, nearly one in five reports made to Call It Out referred specifically to the Voice or the referendum.[31]

## 3 *Phase 3: Hear Me Out*

Hear Me Out extended the model of Call It Out by using artificial intelligence to help individuals access complaint-making systems while retaining the capacity to learn from the complaints that users make. Developed by the NJP in partnership with the UNSW Kaldor Centre for International Refugee Law, K&L Gates and a digital design agency, with support from the NSW Access to Justice Innovation Fund, *Hear Me Out* was launched in NSW on 21 March 2025 as Australia's, if not the world's, first AI-powered complaint platform.[32] It was introduced into the State of Victoria in November 2025 and is currently being Beta tested for Western Australia, with a potential further expansion across all Australian states and territories, and New Zealand.

The platform allows users to describe their experiences in plain language and receive guidance about where and how to make a complaint. At launch, it provided legal resources, factsheets and guidance covering more than 75 complaint bodies across NSW and federal jurisdictions.[33] Hear Me Out processes these plain-language narrative inputs, matches grievances against appropriate oversight entities across various statutory frameworks, and assists users in drafting formal submissions via an automated Complaint Writing Tool.[34] The platform also functions as data infrastructure, by allowing individual complaint assistance and collective evidence-building to occur through the same system.

By capturing structured and de-identified information from complaints, Hear Me Out can identify patterns across institutions and complaint pathways that would be difficult to see from individual matters alone. During its six-month New South Wales pilot, Hear Me Out assisted with more than 400 complaints, with 33 per cent involving police, legal, human rights and discrimination issues.[35] The platform's development subsequently expanded from identifying appropriate complaint pathways to assisting users with the preparation of complaints themselves.

From March 2026 to September 2026, 20,874 people visited the website across 24,895 visits, averaging approximately 136 visits per day. Over the same period, user conversion rates increased from 227 to 1,351 per month, representing an almost six-fold increase. The way users navigate Hear Me Out also demonstrates that the platform is being used to address specific legal and administrative problems. NSW Fair Trading, eSafety, the Health Care Complaints Commission, Fair Work and the Department of Education were among the ten most-viewed pages and together accounted for 13% of all page views. The 26% regular visitor rate also indicates that approximately one in four visitors returns to Hear Me Out each week, demonstrating that the platform is being used repeatedly to formulate complaints, rather than as a one-off source of information.

Unlike rules-based decision trees that require users to self-diagnose their legal problems, Hear Me Out uses retrieval-augmented generation (RAG) architectures to identify relevant information in plain language complaints, and combine that with standardised jurisdictional templates to accurately determine relevant complaint pathways.[36] The significance of this is that Hear Me Out can triage a pathway, provide information about relevant complaint bodies and later, assist with the work required for individuals to pursue that pathway. This reduces the need for users to identify the legal or administrative category into which their problem falls before receiving assistance. Further, integrating data capture into the reporting process helps ensure that individual self-help and collective advocacy are not mutually exclusive, and assists them to consolidate their concerns into a sensible complaint. By designing the underlying artificial intelligence pipeline to capture and structure metadata from every complaint lodged, Hear Me Out functions simultaneously as a legal assistant for the individual and as data infrastructure for civil society.[37] By mapping complaints across more than 300 oversight bodies, Hear Me Out can generate empirical proof of systemic failures, equipping advocacy organisations with the evidence needed to demand institutional accountability and statutory reform.[38]

---

[28] The authors gratefully acknowledge Daniel Ghezelbash for his review of this discussion.

[29] Joint Standing Committee on Aboriginal and Torres Strait Islander Affairs, Parliament of Australia, *Inquiry into Racism, Hate and Violence directed at Aboriginal and Torres Strait Islander people* (2026); National Justice Project, 'What we're calling for in the First Nations racism inquiry' (17 July 2026).

[30] Ibid.

[31] Jumbunna Institute for Indigenous Education and Research and National Justice Project, *If You Don't Think Racism Exists Come Take a Walk with Us: Call It Out Annual Report 2023–2024* (Report, 2025) 7–8; University of Technology Sydney, '"Brave and compelling narratives" captured in racism report' (6 March 2025).

[32] National Justice Project, 'Launch of Australia's first AI-powered complaints platform' (21 March 2025); University of New South Wales, 'A new era for access to justice: AI-powered platform "Hear Me Out" launches in Australia' (24 March 2025).

[33] National Justice Project, 'Launch of Australia's first AI-powered complaints platform' (21 March 2025); University of New South Wales, 'Artificial Intelligence (AI) and disruptive technology' (Web Page).

[34] Dean Moutopoulos et al, 'AI-Powered Platforms for Access to Justice: The Case of Hear Me Out' (UNSW Law Research No 25-13, 2025) 2–4.

[35] National Justice Project, 'Launch of Australia's first AI-powered complaints platform' (21 March 2025); University of New South Wales, 'A new era for access to justice: AI-powered platform "Hear Me Out" launches in Australia' (24 March 2025).

[36] Dean Moutopoulos et al, 'AI-Powered Platforms for Access to Justice: The Case of Hear Me Out' (UNSW Law Research No 25-13, 2025) 2–4.

## IV CHALLENGES AND FUTURE DIRECTIONS

The development of the NJP's technologies has identified areas where the model can be strengthened, particularly in relation to eliciting the relevant facts and circumstances, data governance, long-term funding and access to human assistance. Hear Me Out has been developed with significant attention to the accuracy of its RAG-assisted triage. An ongoing consideration is the accuracy and completeness of unguided user submissions. Moving forward, the objective is to improve the performance of the platform to further prompt users before triaging the appropriate complaint pathway, and then assisting them to make a more detailed complaint.[39]

Aggregating sensitive reports also requires close attention to the legal implications, privacy and data governance standards. These concerns are particularly important for First Nations communities. NJP addresses these concerns by limiting the information collected for example, users are instructed not to provide personal details and neither platform retains identifiable information. For Call It Out, the data is held solely by the Jumbunna Institute, ensuring that information about First Nations experiences remains under First Nations control.[40] Sustainable funding is also necessary for the continued operation and development of these platforms. In September 2026, the Australian Federal Government committed up to $2 million to Call It Out over two years. The funding will secure the continued operation of the Register to incorporate AI tools and the development of a live online data dashboard to provide timely information on which to base anti-racism policy and responses.[41] A further area for development is to add human assistance to its online offerings. Research by the Law and Justice Foundation of NSW (2019) found that 63.8% of legal problems for which respondents used self-help resources also involved the use of a formal legal adviser.[42] This suggests there will be an ongoing need for human assistance in conjunction with digital self-help tools. There is potential for the NJP to build on this model by creating a referral pathway from digital reporting to human assistance. Going forward, the NJP has plans to build a human pathway back into the model, through a dedicated clinic or referral mechanism,[43] connecting digital self-help with personal and systemic advocacy and legal assistance where appropriate and has already undertaken a pilot with Monash University Social Justice Clinic.[44]

## V CONCLUSION

Over six years, the NJP has developed a number of digital tools to address the limitations in conventional complaint and legal assistance systems. Digital reporting has the potential to make Australia's fragmented complaints and oversight landscape more accessible to users, particularly within a milieu where the technology frontier continues expanding, and where individuals may be unfamiliar with the institutions responsible for addressing their complaints or the process required to make a complaint. At the same time, aggregated reporting can reveal patterns that are difficult to identify from individual complaints; the NJP has effectively used this evidentiary base to support parliamentary inquiries, law reform, and strategic litigation. In this way, the NJP's model demonstrates how technology, when thoughtfully designed and responsibly governed, can expand access to justice. The future of justice technology therefore lies in connecting digital self-help with the human and institutional pathways required to act on what reporting reveals. Justice technology is at its most powerful when the act of reporting becomes a means of documenting harm, building evidence and demanding a response.

---

[37] National Justice Project, 'Launch of Australia's first AI-powered complaints platform' (21 March 2025); National Justice Project, *Impact Report 2015–2025* (2025) 26.

[38] Ibid. See also Dean Moutopoulos, Daniel Ghezelbash, Ariane Dozer and George Newhouse, 'How Pro Bono Partnerships Power Hear Me Out' (June 2025) *Pro Bono Voco* (Australian Pro Bono Centre).

[39] Ibid.

[40] National Justice Project, *Impact Report 2015–2025* (2025) 26; University of Technology Sydney, 'Call it Out Racism Register' (Web Page).

[41] University of Technology Sydney, 'Racism register wins federal government backing' (3 September 2026).

[42] Hugh M McDonald, Suzie Forell and Zhigang Wei, 'Uptake of legal self-help resources: what works, for whom and for what?' (2019) 30 *Justice Issues*, Law and Justice Foundation of NSW, 7–8.

[43] National Justice Project, *Impact Report 2015–2025* (2025) 26.

[44] Monash University, 'National Justice Project (LAW4811)', Faculty of Law (Web Page).

## AI USE STATEMENT

Generative AI tools (ChatGPT) were used to help summarise background research materials. All analysis, arguments and conclusions are the authors' own. The authors reviewed and verified all AI-assisted content and take full responsibility for the paper.